\documentclass [amssymb,amsmath,onecolumn, showpacs]{revtex4-1} % twocolumn, 
\usepackage{graphicx,epsfig,amsfonts,amssymb}
\usepackage{bm}% bold math
\usepackage{times}
\usepackage{lipsum}

\newcommand{\beq}{\begin{equation}}
\newcommand{\eeq}{\end{equation}}
\newcommand{\bes}{\begin{subequations}}
\newcommand{\ees}{\end{subequations}}
\newcommand{\bea}{\begin{eqnarray}}
\newcommand{\eea}{\end{eqnarray}}
\newcommand{\ba}{\begin{array}}
\newcommand{\ea}{\end{array}}
\newcommand{\beqn}{\begin{eqnarray*}}
\newcommand{\eeqn}{\end{eqnarray*}}

\begin{document}

\title{Exact transition probabilities in a six-state Landau-Zener system with path interference}

\author {N.~A. {Sinitsyn}$^{a}$ }
\address{$^a$ Theoretical Division, Los Alamos National Laboratory, Los Alamos, NM 87545,  USA}

\begin{abstract}
We identify a nontrivial multistate Landau-Zener model for which transition probabilities between any pair of diabatic states can be determined analytically and exactly.
In the semiclassical picture, this model features the possibility of interference of different trajectories that connect the same initial and final states. Hence, transition probabilities are generally not  described by the incoherent successive application of the Landau-Zener formula. We discuss reasons for integrability of this system and provide numerical tests of the suggested expression for the transition probability matrix.   
\end{abstract}
\date{\today}

\maketitle

%\section{Introduction}
 
%Explicitly time-dependent Schr\"odinger equation is
%of importance for numerous applications. 
%Quantum mechanical nonadiabatic transitions have been studied in molecular collisions for long time \cite{rozen, nikitin, osherov}.
%Relatively recently, 

%Multichannel nonadiabatic processes are frequently found in control of quantum many-body states in electronics, magnetic systems, and Bose condensates \cite{app-el,app-bose,app-spin,app-exp}.  
 %Condensed matter physicists often achieve an insight about their systems by finding a proper model from a large variety of exactly solvable systems with well defined thermodynamic limit. Unfortunately, most of such known models correspond to the  stationary  Schr\"odinger equation. Exact solutions of explicitly time-dependent systems with a large number of states are  rare. 
 %They are also often trivially reducible to independent oscillators or non-interacting spins. 
 %It becomes important for the future progress of this field to explore new integrable models of quantum mechanical evolution with time-dependent parameters and with a large, possibly macroscopic, number of interacting states. 
 %%%%%%%%%%%%%%%%%%%%%%%%%%%%%%%%%%%%%%%%%%%%%%%%%%%%%%%%%%%%%%%%%%%%%%%%%%%%%%%%%%%%%%%%%%%
 
% \cite{book,maj,landau, LZ, stuck}

The multistate version of the two-state Landau-Zener model is one of the most fundamental systems in nonstationary quantum mechanics \cite{book, be}, which finds numerous applications in atomic and condensed matter physics  \cite{recent}.
It considers the evolution of $N$  states described by the Sch\"odinger equation with parameters that change linearly with time:
\begin{equation}
i\frac{d\psi}{d t} = \left(\hat{A} +\hat{B}t  \right)\psi.
\label{mlz}
\end{equation} 
Here, $\psi$ is the state vector in a space of $N$ states; $\hat{A}$ and $\hat{B}$ are constant Hermitian $N\times N$ matrices.  One can always choose the, so-called, {\it diabatic basis} in which the matrix $\hat{B}$ is diagonal,
and if any pair of its elements are degenerate then the corresponding off-diagonal element of the matrix $\hat{A}$ can be set to zero, that is
\beq
B_{ij}= \delta_{ij}\beta_i, \quad  A_{nm}=0\,\,\, {\rm if} \,\, \beta_{n}=\beta_{m},\quad n\ne m \in (1,\ldots N).
\label{diab1}
\eeq
Constant parameters $\beta_i$  are called the {\it slopes of diabatic levels};  nonzero off-diagonal elements of the matrix $\hat{A}$ in the diabatic basis are called the {\it coupling constants},  and the diagonal elements of the Hamiltonian, $H_{ii}=\beta_i t +\epsilon_i$, where $\epsilon_i \equiv A_{ii}$, are called the {\it diabatic energies} \cite{sinitsyn-14pra}. 

The goal of the multistate Landau-Zener theory is to find the scattering $N\times N$ matrix $\hat{S}$, whose element $S_{nn'}$ is the amplitude of the diabatic state $n$ at $t  \rightarrow +\infty$, given that at $t \rightarrow -\infty$ the system was in the $n'$-th diabatic state. In most cases, only the related matrix $\hat{P}$, $P_{n' \rightarrow n}=|S_{nn'}|^2$,  called the {\it matrix of transition probabilities}, is of interest. 
Generally, for $N>2$, the analytical solution of the model (\ref{mlz}) is unknown. Nevertheless, a number of exactly solvable cases with specific forms of matrices $\hat{A}$ and $\hat{B}$  have been derived \cite{no-go,mlz-1,do,multiparticle, bow-tie-path, reducible,bow-tie,chain}.

Several influential results in the multistate Landau-Zener theory have been originally proposed as conjectures  based on intuition and numerical calculations. This includes the solution of the generalized bow-tie model \cite{bow-tie-path}, Brundobler-Elser formula \cite{be}, and the no-go rule \cite{no-go}. The subsequent search for the proofs  \cite{mlz-1} not only confirmed these conjectures but also advanced our understanding of solvable systems in nonstationary quantum mechanics  beyond the multistate Landau-Zener model \cite{sinitsyn-14pra,armen}.

In this article, we introduce one more  system of the type (\ref{mlz}), which matrix of transition probabilities can be written explicitly. This model describes the system of $N=6$ interacting states with the Hamiltonian 
\begin{equation}
\hat{H}=\left( 
\begin{array}{cccccc}
\beta t +\epsilon_1 & 0 & 0 & 0 & -g_1 & -g_2 \\
0 & \beta t & 0 & g_1 & 0 & -g_3  \\
0 & 0 & \beta t-\epsilon_2 & g_2 & g_3 & 0 \\
0&g_1& g_2& \epsilon_1& 0&0 \\
-g_1 & 0& g_3 & 0& 0& 0 \\
-g_2& -g_3& 0& 0& 0& -\epsilon_2
\end{array}
\right),
\label{ham1}
\end{equation} 
where parameters $\epsilon_{1}$, $\epsilon_2$, $g_1$, $g_2$, $g_3$, $\beta$ are arbitratry constants and $t$ is time. 

\begin{figure}%[!htb]
\scalebox{0.33}[0.33]{\includegraphics{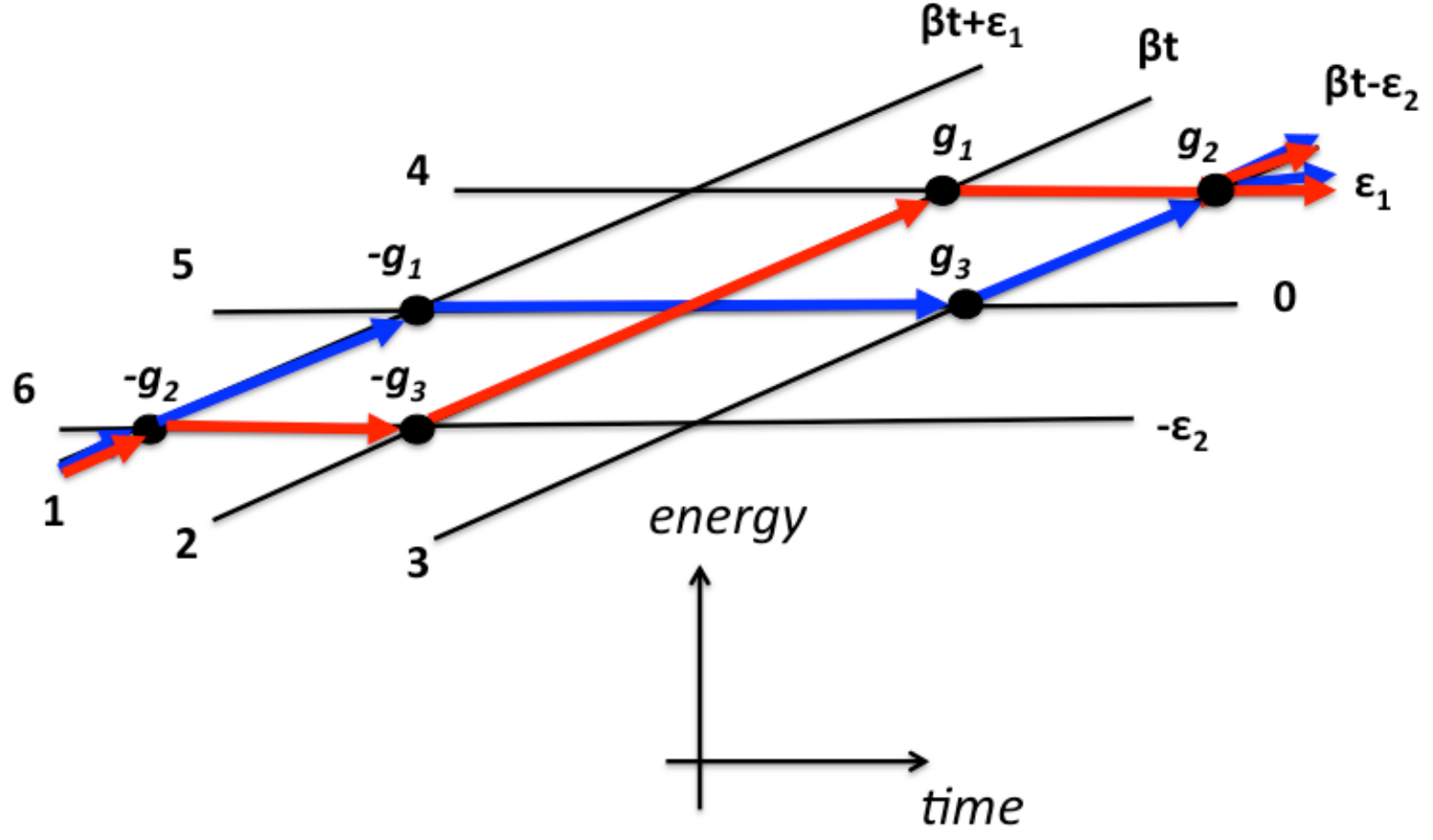}}
\hspace{-2mm}\vspace{-4mm}   
\caption{ (Color online) Diabatic energy levels (straight black lines) and their pairwise couplings (parameters $\pm g_i$, $i=1,2,3$) shown at corresponding level intersections. Red and Blue arrows illustrate trajectories that start at the level-1 and end either at the level-4 or at the level-3 with, respectively, destructive and constructive interference.}
\label{levels}
\end{figure}

It is convenient to illustrate parameters of this model on a diagram that plots diabatic energies as functions of time and marks the pairwise intersections of diabatic levels by corresponding coupling constants, as shown in Fig.~\ref{levels}. For the Hamiltonian (\ref{ham1}), this diagram shows that the model describes two crossing bands of energy levels, with three levels in each band.

%The accumulated knowledge about exactly solvable cases of Eq.~(\ref{mlz})  has been recently used to address questions that reach properties of far more general nonstationary quantum systems \cite{routers}.
%%%%%%%%%%%%%%%%%%%%%%%%%%%%%%%%%%%%%%%%%%%%%%%%%%%%%%%%%%%%%%%%%%%%%%%%%%%%%%%%%%%%%%%%%%%

When diabatic energy levels in each band are well separated from each other, i.e. $\epsilon_1,\epsilon_2 \gg g_1,g_2,g_3$, one can justify the semiclassical approach to estimate transition amplitudes. During time intervals for which all levels are well separated, diabatic states are almost eigenstates of the Hamiltonian. The
adiabatic theorem predicts then that transitions between such states are exponentially suppressed. The adiabaticity condition is violated only for pairs of crossing diabatic states. After passing through such a crossing point, a system can remain on the initial level or turn to another one. Hence, one can visualize possible trajectories of the system as we illustrate in Fig.~\ref{levels}. The blue and the red color arrows indicate the semiclassically allowed trajectories that connect the initially populated state at the level-1 with the levels 3 and 4. Here we note that  only trajectories that turn to the right or upward are allowed in Fig.~\ref{levels} in the semiclassical limit.
% The latter property excludes the possibility of loop-like trajectories, which are usually responsible for localization effects.

The transition probability between a pair of states after a passage through a crossing point is  described by the two-state Landau-Zener (LZ) formula. In our case, if a diabatic level crosses another one with a corresponding nonzero coupling $g_j$, $j\in (1,2,3)$, then with a probability  
\beq
p_j=e^{-2\pi |g_j|^2|/\beta}
\label{lz1}
\eeq
 the system will remain in the same diabatic state after the level crossing. Respectively, the probability to turn to another level is $q_j=1-p_j$.
%%%%%%%%%%%%%%%%%%%%%%%%%%%%%%%%%%%%%%%%%%%%%%%%%%%%%%%%%%%%%%%%%%%%%%%%%%%%%%%%%%%%%%%%%%%
\begin{figure}%[!htb]
\scalebox{0.45}[0.45]{\includegraphics{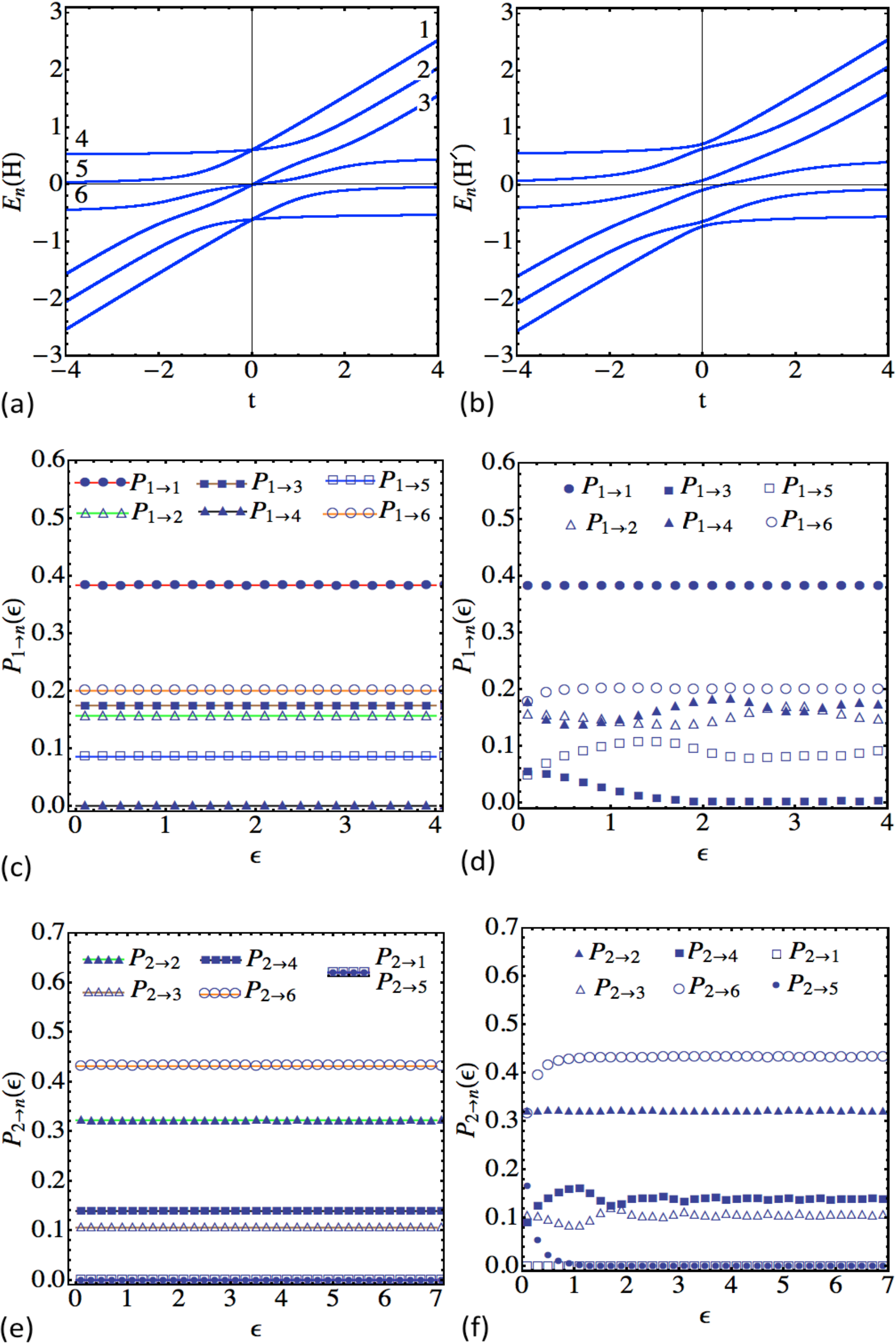}}  %energy1.pdf
\hspace{-2mm}\vspace{-4mm}   
\caption{(Color online) (a-b) Typical dependence of adiabatic energies (eigenvalues of the Hamiltonian matrix) on time $t$. The model in (a) corresponds to the Hamiltonian~(\ref{ham1}) and (b) corresponds to the Hamiltonian~(\ref{ham2}).
The case (a) features three exact level crossings at $t=0$ while these degeneracies are lifted in the case (b). 
(c-d) Transition probabilities from the level-1 as functions of the level separation for $\epsilon_1=\epsilon$, $\epsilon_2=1.5\epsilon$, $g_1=0.25$, $g_2=0.3$, $g_3=0.35$, $\beta=1$. The case (c) corresponds to the model (a)   and (d) corresponds to the model (b).
(e-f) Transition probabilities from the level-2 as functions of the level separation for $g_1=g_2=g_3=0.3$, $\epsilon_1=\epsilon_2=\epsilon$, $\beta=1$. The case (e) corresponds to (a), and (f) corresponds to  (b).
All discrete points correspond to results of direct numerical simulations of the evolution  from $t=-1000$ to $t=1000$, and solid lines are theoretical predictions of Eq.~(\ref{prob1}). Details of the numerical algorithm are discussed in the supplementary material for Ref.~\cite{sinitsyn-13prl}.
}
\label{energy}
\end{figure}
%%%%%%%%%%%%%%%%%%%%%%%%%%%%%%%%%%%%%%%%%%%%%%%%%%%%%%%%%%%%%%%%%%%%%%%%%%%%%%%%%%%%%%%%%%%

By applying the LZ-formula to pairs of diabatic level crossings according to their appearance in a chronological order one can estimate the amplitude of a semiclassical trajectory.  For example, consider the process in Fig.~\ref{levels} that starts at the level-2 and ends at the level-4.  Semiclassically, there is only a single trajectory that connects these states. It starts at the level-2, passes through the intersection with the level-6 without changing the diabatic state, then it turns to the level-4 at the corresponding intersection and then goes through the intersection with the level-3 without changing the level. The amplitude of such a process is
given by
\beq
S_{42} = \sqrt{p_3(1-p_1)p_2} e^{i(\phi_d^{42}+\phi_{LZ}^{42})},
\label{tra1}
\eeq
where $\phi_d$ is the dynamic phase of a trajectory: 
\beq
\phi_d = -\int \, (\beta_{k(t)}  t + \epsilon_{k(t)})  \, dt, 
\label{pd}
\eeq
and $\phi_{LZ}$ is the ``LZ-phase" that the system accumulates after transitions through the crossing points. Time-dependence of indexes in (\ref{pd}) indicates that different parts of a trajectory generally correspond to system's presence at levels with different values of the index $k$ and consequently different values of parameters $\beta_k\in (0,\beta)$ and $\epsilon_k \in (-\epsilon_2,0,\epsilon_1)$.  Since there is only a single trajectory connecting levels 4 and 2, the phase factor does not influence the corresponding transition probability: 
\beq
P_{2\rightarrow 4}= p_3(1-p_1)p_2.
\label{p42}
\eeq

The situation becomes more complex when there are more than one trajectory connecting initial and  final states. In our case, such trajectories start from either the level 1 or 6 and end at either the level 3 or 4. In an arbitrary nonintegrable model, relative dynamic phases of different trajectories change violently with changing parameters. However, the model with the Hamiltonian (\ref{ham1}) is tuned so that different trajectories connecting the same initial and final states acquire the same dynamic phase. Consider, e.g. transitions from the level-1 to the level-4. One can  find that the blue-color trajectory in Fig.~\ref{levels} spends the same amount of time at the level-1 with the constant part of the diabatic energy equal to $\epsilon_1$ as the red trajectory spends at the level-4 with the same value of the constant part of the diabatic energy, etc. 
Another way to see that the dynamic phase is the same for both paths is to use the property that the dynamic phase difference between two trajectories is proportional to the area between them  \cite{bow-tie-path}. In  Fig.~\ref{levels}, the blue and the red trajectories enclose two plaquettes of equal size but different sign.  
Hence, despite the trajectory amplitude  interference, the dynamic phase drops out of the final expression for the transition probabilities in the model (\ref{ham1}). 

The role of the LZ-phase in the independent crossing approximation was discussed by Demkov and Ostrovsky in  \cite{bow-tie-path}. They showed that the effect of this phase can be trivially included by assuming that, at each pairwise crossing point, if a system remains at the initial diabatic level it does not gain an additional phase factor but, if it changes the diabatic state, the amplitude of the trajectory acquires an additional factor $i$.  We found that this rule is not sufficient for our case because it was derived for real positive coupling constants. In the model in  Fig.~\ref{levels}, all  coupling constants appear with opposite sings at different intersection points. To account for this case, consider a simple Landau-Zener system:
\beq
i\frac{d}{dt} a=\beta_1 t a +|g|e^{i\phi_g} b,\quad i\frac{d}{dt} b=\beta_2 t  b +|g|e^{-i\phi_g} a,
\label{lz1}
\eeq
where we explicitly separated the absolute value and the phase of the coupling constant. By a change of variables, $b \rightarrow be^{-i\phi_g}$,  the system (\ref{lz1}) transforms into the one with a real positive coupling:
\beq
i\frac{d}{dt} a=\beta_1 t a +|g| b,\quad i\frac{d}{dt} b=\beta_2 t  b +|g| a.
\label{lz2}
\eeq
This means that the transition amplitude from the diabatic state described by the amplitude $a$ into the other diabatic state is different in models (\ref{lz1}) and (\ref{lz2}) by a phase factor $e^{-i\phi_g}$. In application to the multistate Landau-Zener models, this means that each time a semiclassical trajectory switches between different diabatic levels, its amplitude acquires a phase factor $ie^{-i\phi_g}$, were $\phi_g$ is the phase of the coupling constant at a corresponding level intersection.  Following this rule, we should assume for the model in Fig.~\ref{levels} that whenever a trajectory changes levels, it acquires a phase factor $\pm i$ for, respectively, positive and negative values of the couplings at intersections.

Applying this rule to the transition from the level-1 to the level-4 we find two contributing amplitudes:
\bea
S^{\rm red}_{41} =  (-1)^2i^3 \sqrt{(1-p_2)(1-p_3)(1-p_1)p_2}, \quad
S^{\rm blue}_{41} =  -i^3\sqrt{p_2(1-p_1)(1-p_3)(1-p_2)}.
\label{ampt}
\eea
The sum of those amplitudes is identically zero, i.e., trajectories interfere destructively so that the transition from the level-1 to the level-4 is forbidden: $P_{1\rightarrow 4}=0$. 
For the transition from the level-1 to the level-3 we find: 
\bea S^{\rm red}_{31} = (-1)^2i^4 \sqrt{(1-p_2)(1-p_3)(1-p_1)(1-p_2)} ,\quad
S^{\rm blue}_{31} =- i^2 \sqrt{p_2(1-p_1)(1-p_3)p_2} .
\label{ampt2}
\eea
Amplitudes (\ref{ampt2}) interfere constructively, so that $P_{1 \rightarrow 3}=(1-p_3)(1-p_1)$. We are now in a position to summarize the prediction of the independent crossing approximation for the form of the transition probability matrix: 
\beq
\hat{P} = \left( 
\begin{array}{cccccc}
p_2p_1&q_2q_3 p_1 & q_1q_3 & 0 & p_2q_1p_3 & q_2p_3 \\
0 & p_3 p_1 & p_3q_1q_2 &p_3q_1p_2 & 0 & q_3  \\
0 & 0 & p_3p_2 & p_3q_2 & q_3 & 0 \\
0&q_1& p_1q_2& p_1p_2& 0&0 \\
q_1 & 0& p_1q_3 p_2& p_1q_3q_2& p_1p_3& 0 \\
q_2p_1& p_2q_3p_1& 0& q_1q_3& q_2q_1p_3&p_2p_3
\end{array}
\right), \quad q_n\equiv 1-p_n.
\label{prob1}
\eeq

So far, our derivation of the transition probability matrix was purely semiclassical. It is expected to be predictive only for sufficiently large values of parameters $\epsilon_1$ and $\epsilon_2$. In what follows, we are going to claim that for the model with the Hamiltonian (\ref{ham1}) the matrix (\ref{prob1}) is actually exact, i.e. it is valid for arbitrary values of all parameters.

In order to understand what is special about the model (\ref{ham1}), it is instructive to compare this model with a similar one that has the Hamiltonian 
\begin{equation}
\hat{H}'=\left( 
\begin{array}{cccccc}
\beta t +\epsilon_1 & 0 & 0 & 0 & g_1 & g_2 \\
0 & \beta t & 0 & g_1 & 0 & g_3  \\
0 & 0 & \beta t-\epsilon_2 & g_2 & g_3 & 0 \\
0&g_1& g_2& \epsilon_1& 0&0 \\
g_1 & 0& g_3 & 0& 0& 0 \\
g_2& g_3& 0& 0& 0& -\epsilon_2
\end{array}
\right),
\label{ham2}
\end{equation} 
which is different from (\ref{ham1}) only by removal of the minus signs in front of some coupling constants. Our discussion about interference and cancellation of the relative dynamic phase equally applies to the model (\ref{ham2}). For example, transitions that correspond to a single allowed semiclassical path, such as from the level-2 to any other state, should have the same probabilities in both models. 
The only semiclassical difference is that now  constructive and destructive interference correspond to different transitions.

Nevertheless, models (\ref{ham1}) and (\ref{ham2}) show very different behavior beyond the semiclassical limit. The crucial difference between them becomes apparent if we plot eigenvalues of their Hamiltonians (adiabatic energies) as functions of the parameter $t$, as shown in Figs.~\ref{energy}(a,b). Figure~\ref{energy}(a) corresponds to the model (\ref{ham1}). It demonstrates that at $t=0$ the system experiences three pairs of exact crossings of {\it adiabatic} energies. In contrast,  Figure~\ref{energy}(b) shows that, in the model (\ref{ham2}), those pairwise degeneracies are lifted. Returning to the diabatic basis, we find that the moment $t=0$ also corresponds to the crossings between diabatic levels: 1 and 4, 2 and 5, 3 and 6. Such crossing levels are not coupled to each other directly, however, higher order processes can and do induce transitions between them. Nevertheless, the symmetry of the model (\ref{ham1}) is sufficient to completely decouple those pairs of states at $t=0$ but in the model (\ref{ham2}) mini-gaps open up. Such mini-gaps are the signatures of the breakdown of the semiclassical approximation.

Let us compare now the transition probabilities in models (\ref{ham1}) and (\ref{ham2}) as functions of the level separation, as we illustrate in Figs.~\ref{energy}(c-f) for different initial conditions and different ratios of $\epsilon_1/\epsilon_2$. In the case of the model (\ref{ham1}), numerically obtained transition probabilities are found to be independent of the values of the parameters $\epsilon_1$ and $\epsilon_2$, and they perfectly agree with the theoretical prediction (\ref{prob1}).   In contrast, in the model (\ref{ham2}), transition probabilities generally depend on  $\epsilon_1$, and $\epsilon_2$ nontrivially and saturate only in the limit $\epsilon_1,\epsilon_2 \gg g_1,g_2,g_3$, as it is expected from the semiclassical approximation. 

%%%%%%%%%%%%%%%%%%%%%%%%%%%%%%%%%%%%%%%%%%%%%%%%%%%%%%%%%%%%%%%%%%%%%%%%%%%%%%%%%%%%%%%%%%%
\begin{figure}%[!htb]
\scalebox{0.65}[0.65]{\includegraphics{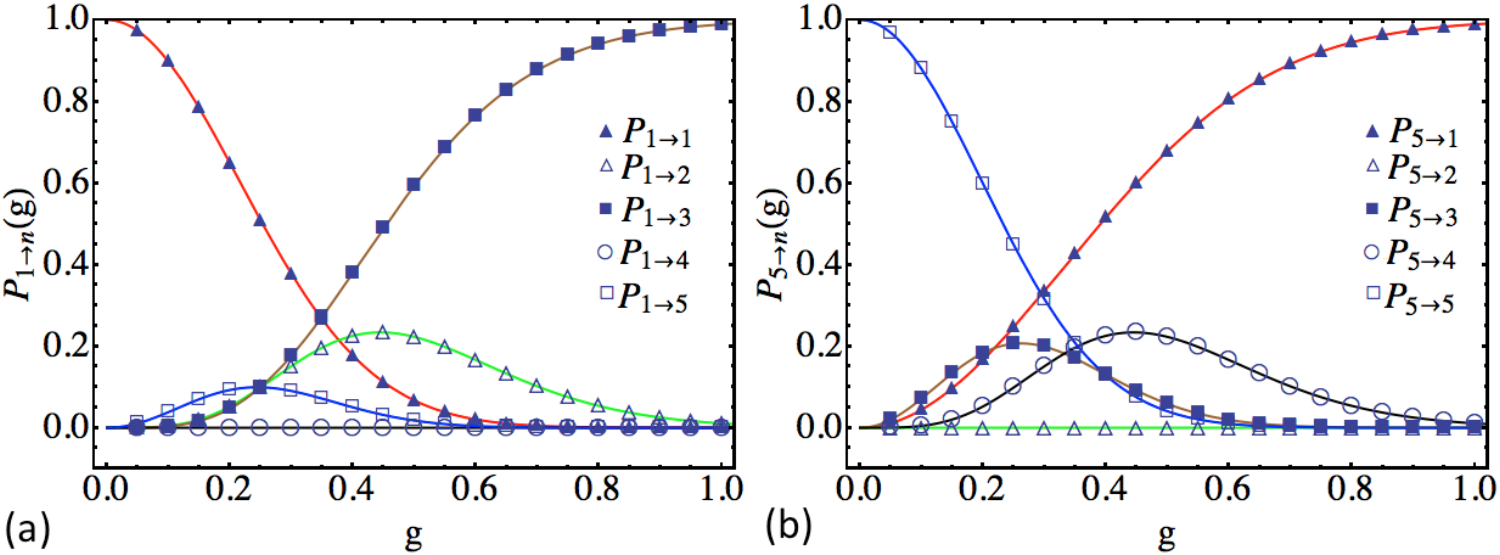}}
\hspace{-2mm}\vspace{-4mm}   
\caption{ (Color online) Numerical test of Eq.~(\ref{prob1}) for initially populated (a) level-1 and (b) level-5.
All discrete points correspond to results of direct numerical simulations of the evolution with the Hamiltonian (\ref{ham1}) from $t=-1000$ to $t=1000$. Solid lines are theoretical predictions of Eq.~(\ref{prob1}). The choice of parameters is: 
$\epsilon_1=0.25$, $\epsilon_2=0.35$, $g_1=0.85g$, $g_2=1.0g$, $g_3=1.15g$, $\beta=1$.}
\label{coupling}
\end{figure}
%%%%%%%%%%%%%%%%%%%%%%%%%%%%%%%%%%%%%%%%%%%%%%%%%%%%%%%%%%%%%%%%%%%%%%%%%%%%%%%%%%%%%%%%%%

Comparison of the models (\ref{ham1}) and (\ref{ham2}) does support the conjecture that the transition probability matrix (\ref{prob1}) is the exact nonperturbative solution of the multistate Landau-Zener problem with the Hamiltonian (\ref{ham1}). As an additional numerical proof, we provide Fig.~\ref{coupling} that compares the theoretical prediction (\ref{prob1}) with numerical simulations of the model (\ref{ham1}) with different choices of the coupling constants. All our numerical tests have shown perfect agreement with Eq.~(\ref{prob1}).

{\it In conclusion}, we identified and explored the Landau-Zener-like model of two crossing bands, for which we determined the matrix of transition probabilities. This model shows a relatively complex behavior due to the possibility of semiclassical path interference leading to either constructive or destructive interference.  Our numerical simulations confirm that the transition probability matrix, which we derived in a semiclassical framework, is exact i.e. it describes an arbitrary choice of the model parameters.  

Our model was identified on the basis of its similarity with other  known exactly solvable systems, such as  the reducible 6-state model \cite{multiparticle} and the generalized bow-tie model \cite{bow-tie-path}. For such previously solved systems, the semiclassically derived and exact expressions for transition probabilities coincide. Two  properties are also shared  by our model with these systems: (i) the absence of the dynamic phase effect on transition probabilities in the semiclassical framework and (ii) the exact crossing of adiabatic energies at intersections of diabatic states without direct couplings. 
This finding suggests that our model and the previously solved systems \cite{multiparticle,bow-tie-path}  are likely the instances of a more general class of solvable models that share the same properties. It should be interesting to identify new candidate systems of the type of Eq.~(\ref{mlz}) with properties (i-ii) and test them for integrability. 
Further progress in this direction will be also achieved by deriving a mathematically rigorous proof of Eq.~(\ref{prob1}).    

{\it Acknowledgment}. The work
was carried out under the auspices of the National Nuclear
Security Administration of the U.S. Department of Energy at Los
Alamos National Laboratory under Contract No. DE-AC52-06NA25396. Author also thanks the support from the LDRD program at LANL.


\begin{thebibliography}{100}


\bibitem{be}S. Brundobler and V. Elser, J. Phys. A {\bf 26}, 1211 (1993)


\bibitem{book} H. Nakamura, ``Nonadiabatic Transition", World Scientific Publishing Company, 2nd Edition (2012); M. S. Child, ``Molecular Collision Theory", Dover Publications (2010); E. E. Nikitin, S. Y. Unanskii, ``Theory of Slow Atomic Collisions" , Springer Series in Chemical Physics (Book 30), Springer (2011)
%\bibitem{maj} E. Majorana, Nuovo Cimento {\bf 9} (2), 43 (1932)
%\bibitem{landau}  L. D. Landau, Physik Z. Sowjetunion {\bf 2}, 46 (1932)
%\bibitem{LZ} C. Zener, Proc. R. Soc. A {\bf 137}, 696 (1932)
%\bibitem{stuck} E. C. G. St\"uckelberg, Helv. Phys. Acta {\bf 5}, 369 (1932)

\bibitem{recent} Y-A. Chen, S. D. Huber,  S. Trotzky,	 I. Bloch, and E. Altman, Nature Phys. 7, 61 (2011); C. Kasztelan, S. Trotzky, Y.-A. Chen, I. Bloch, I. P. McCulloch, U. Schollw\"ock, and G. Orso, Phys. Rev. Lett. 106, 155302 (2011); S F. Caballero-Benitez and R. Paredes, Phys. Rev. A 85, 023605 (2012); W. Tschischik, M. Haque, R. Moessner, Phys. Rev. A 86, 063633 (2012)

\bibitem{sinitsyn-14pra} N. A. Sinitsyn, Phys. Rev. A {\bf 90}, 062509 (2014)



%\bibitem{shytov}  A. V. Shytov, Phys. Rev. A {\bf 70}, 052708 (2004)
\bibitem{no-go} N. A. Sinitsyn,  J. Phys. A {\bf 37} (44), 10691 (2004)

\bibitem{mlz-1}   A. V. Shytov, Phys. Rev. A {\bf 70}, 052708 (2004);   M. V. Volkov and V. N. Ostrovsky, J. Phys. B: At. Mol. Opt. Phys. {\bf 37}, 4069 (2004); M. V. Volkov and V. N. Ostrovsky,  J. Phys. B: At. Mol. Opt. Phys. {\bf 38}, 907 (2005); B. E. Dobrescu and N. A. Sinitsyn, J. Phys. B: At. Mol. Opt. Phys. {\bf 39}, 1253 (2006)


\bibitem{do} Yu. N. Demkov and V. I. Osherov, Zh. Exp. Teor. Fiz. {\bf 53}, 1589 (1967) [Sov. Phys. JETP {\bf 26}, 916 (1968)]; A. A. Rangelov, J. Piilo, and N. V. Vitanov, Phys. Rev. A {\bf 72}, 053404 (2005)

\bibitem{multiparticle} N. A. Sinitsyn, Phys. Rev. B {\bf 66}, 205303 (2002)
\bibitem{reducible} J. Dziarmaga, Phys. Rev. Lett. {\bf 95}, 245701 (2005);  M. V. Volkov and V. N. Ostrovsky, Phys. Rev. A {\bf 75}, 022105 (2007)

\bibitem{bow-tie-path}  Y. N. Demkov and V. N. Ostrovsky, Phys. Rev. A {\bf 61}, 032705 (2000)
\bibitem{bow-tie} Yu. N. Demkov and V. N. Ostrovsky, J. Phys. B {\bf 28}, 403 (1995); V. N. Ostrovsky and H. Nakamura, J. Phys. A {\bf 30}, 6939 (1997);  Y. N. Demkov and V. N. Ostrovsky, J. Phys. B {\bf 34}, 2419 (2001); C. E. Carroll and F. T. Hioe, J. Phys. A {\bf 19}, 1151 (1986)

\bibitem{chain}  N. A. Sinitsyn, Phys. Rev. A {\bf 87}, 032701 (2013); V. L. Pokrovsky and N. A. Sinitsyn, Phys. Rev. B {\bf 65}, 153105 (2002)


\bibitem{armen} A. Patra, E. l. A. Yuzbashyan, Preprint arXiv/1412.5880 (2014)

\bibitem{sinitsyn-13prl} N. A. Sinitsyn, Phys. Rev. Lett. {\bf 110}, 150603 (2013)

\end{thebibliography}
\end{document}